\documentclass[lettersize,journal]{IEEEtran}
\usepackage{booktabs, makecell, multirow}

\usepackage{siunitx}
\usepackage{amsmath,amsfonts}
\usepackage{algorithmic}
\usepackage{algorithm}
\usepackage{array}
\usepackage[caption=false,font=normalsize,labelfont=sf,textfont=sf]{subfig}
\usepackage{textcomp}
\usepackage{stfloats}
\usepackage{url}
\usepackage{verbatim}
\usepackage{graphicx}
\usepackage{cite}
\usepackage{subcaption,lipsum}
\usepackage{threeparttable}
\begin{document}
\title{Meta-learning-based percussion transcription and $t\bar{a}la$ identification from low-resource audio}
\author{Rahul Bapusaheb Kodag, Vipul Arora ~\IEEEmembership{Senior Member,~IEEE,}
\thanks{The work was financially supported by Prasar Bharati, the public broadcasting corporation of India.}}



\maketitle

\begin{abstract}
This study introduces a meta-learning-based approach for low-resource Tabla Stroke Transcription (TST) and $t\bar{a}la$ identification in Hindustani classical music. Using Model-Agnostic Meta-Learning (MAML), we address the challenges of limited annotated datasets and label heterogeneity, enabling rapid adaptation to new tasks with minimal data.

The method is validated across various datasets, including tabla solo and concert recordings, demonstrating robustness in polyphonic audio scenarios. We propose two novel $t\bar{a}la$ identification techniques based on stroke sequences and rhythmic patterns. Additionally, the approach proves effective for Automatic Drum Transcription (ADT), showcasing its flexibility for Indian and Western percussion music. Experimental results show that the proposed method outperforms existing techniques in low-resource settings, significantly contributing to music transcription and studying musical traditions through computational tools.

\end{abstract}

\section{Introduction}\label{Introduction}
Music Information Retrieval (MIR) focuses on extracting meaningful features from music recordings, enabling diverse applications in music education, production, and musicology \cite{D4}. Among its many tasks, Automatic Music Transcription (AMT) is particularly challenging, as it involves converting audio recordings into symbolic representations such as scores or MIDI files. AMT supports critical tasks such as melody extraction, chord recognition, beat tracking, and rhythm analysis, forming the foundation for advanced music research and applications \cite{D12}. While substantial progress has been made in transcribing melodic instruments such as the piano, guitar, strings, and winds, the transcription of percussive instruments remains relatively under-explored \cite{D4,R6}. Furthermore, research progress in the percussion transcription of Hindustani classical music has been significantly slower compared to Western music \cite{T7}.

\par The tabla is a key instrument in Hindustani music, serving as the rhythmic backbone and accompanying singers and instrumentalists. It helps maintain the structure and essence of the chosen raga (melodic framework). Tabla Stroke Transcription (TST) is crucial for analyzing the complex rhythmic patterns ($t\bar{a}las$) that form the basis of Hindustani music. Accurate TST also aids in preserving and digitizing traditional music and creating learning tools for learners and musicians. Like Automatic Drum Transcription (ADT) in Western music, which focuses on drum sound transcription to understand rhythm structures, organize music collections and develop advanced tools, TST aims to identify and classify rhythmic events. However, unlike ADT, TST presents distinct challenges due to the tabla’s unique physical and timbral characteristics compared to drums \cite {D4, S12}.

\par TST and ADT are challenging due to the complexity of audio signals in music tracks, where multiple instruments (polyphony) often mask rhythmic sounds. In Western music, bass guitars may overlap with bass drums, while melodic instruments or vocals can obscure tabla strokes in Hindustani music. Transcription methods address these challenges at different levels of complexity: Drum/Tabla -only Transcription (DTD) focuses on isolated recordings, avoiding polyphony. Drum/Tabla Transcription with Percussion (DTP) includes additional percussive elements, adding complexity. The most challenging, Drum/Tabla Transcription with Melodic Instruments (DTM), handles full audio mixes with all instruments and vocals. Since DTM represents most real-world musical tracks, it is the most critical and complex task in transcription.

\par Most existing ADT and recent TST methods rely on deep learning (DL) models due to their ability to handle complex audio mixtures \cite{S9, S10}. However, DL models require large, annotated datasets for supervised training, which are scarce. Creating these datasets is challenging because annotations must be precisely aligned with the audio, a labour-intensive process requiring skilled musicians, and many annotated tracks are copyrighted, limiting their public availability. Synthetic data generated from MIDI tracks has been used to produce large-scale training datasets with accurate labels, but such datasets often perform less effectively than real-world annotated data, making dataset scarcity a significant barrier for DL-based ADT and TST.

\par Beyond low-resource conditions, a second challenge is label heterogeneity: stroke labels in tabla and drum datasets differ due to performance styles, labelling schemes, and musical traditions. For example, tabla datasets may follow $ghar\bar{a}n\bar{a}$ conventions, while Western drum datasets use distinct ontologies. These divergences lead to task heterogeneity with mismatched label spaces, reducing the reliability of standard transfer learning approaches that assume shared label ontologies and often requiring time-consuming label harmonization. Model-Agnostic Meta-Learning (MAML) enables fast adaptation across tasks with distinct label spaces \cite{F1, R3, R4, R5}. This paper presents an MAML-based stroke transcription method addressing data scarcity and label inconsistency.

\par In this work, we address the challenge of performing accurate Tabla Stroke Transcription (TST) and Automatic Drum Transcription (ADT) with limited annotated data. Based on the transcribed stroke sequences from TST, we propose methods for $t\bar{a}la$ identification, as $t\bar{a}la$ forms the rhythmic backbone of Hindustani music. Identifying the $t\bar{a}la$ enables a deeper analysis of Hindustani music, and existing music sequence matching approaches \cite{T1,T3} are not directly applicable in this context.

The key contributions of this work are summarized as follows:
\begin{enumerate}
    \item A novel meta-learning-based approach for TST is proposed and validated on both tabla solo and full concert recordings. This method is also extended to ADT tasks using the DTD, DTP, and DTM datasets, demonstrating its generalization across datasets.
    \item Two new $t\bar{a}la$ identification methods based on tabla stroke sequences are introduced.
    \item A synthetic dataset is curated from the existing mridangam stroke dataset.
\end{enumerate}

The rest of the paper is organized as follows: Section \ref{literature} provides an overview of related literature, Section \ref{s_proposed_system} details the proposed method for TST and ADT, Section \ref{Datasets} presents information about the datasets used, Section \ref{Exp_set_up} outlines the experimental setup, Section \ref{results} discusses the experimental evaluation and findings, and Section \ref{conclusion} concludes the paper with key insights and future directions. The dataset and supplementary materials are available at \url{https://github.com/madhavlab/2025_drumtrans_rkodag}

\section{Literature review}\label{literature}
\subsection{TST and ADT} 
Significant work has been done on ADT in Western music \cite{D4}. Recent ADT methods utilize deep learning with various model architectures and feature representations \cite{D1, D3, D4, D5, D7, D10, D11, D12}. ADT identifies stroke onsets, marked by sudden energy increases. The attack-decay-sustain-release (ADSR) sequence forms the general temporal structure of a stroke. Detecting strokes in a drum kit is relatively straightforward due to distinct timbral structures and variations in shape and material \cite{D4}. Drum strokes exhibit sharp attacks, rapid decay, and consistent repetition \cite{S11}. In contrast, TST is more complex due to timbral similarities and variable stroke shapes from the same instrument, necessitating more sophisticated transcription methods and making rhythmic analysis in Hindustani music more intricate. This work focuses on TST, emphasizing $t\bar{a}la$ identification and further validation of TST methods alongside ADT.
\par The literature on TST employs two main approaches: segment-classify and deep learning. The segment-classify method detects stroke onsets using thresholding on spectral flux, then classifies segments with various classifiers. Studies \cite{S1,S2,S3,S4} follow this approach. Early work used Gaussian Mixture Models (GMM) and Hidden Markov Models (HMM), later extended with spectral and temporal features for neural networks and SVM classifiers \cite{S1,S2, S3}. Study \cite{S4} focuses on the acoustic features and ADSR properties of tabla strokes. The DL approach involves spectrograms or Mel spectrograms processed through networks for transcription, with recent studies \cite{S8,S9, S10} employing CNNs, data augmentation, and transfer learning techniques.
\par Early studies \cite{S1, S2, S3, S4} trained models on limited datasets, often from a single tabla. Recent studies \cite{S9, S10} used more realistic datasets with harmonium accompaniment but still employed only four stroke classes, insufficient for actual $t\bar{a}la$ identification. Additionally, these models have not been tested on concert data with vocals and other instruments, raising doubts about their performance on actual stroke classes and concert datasets.
\subsection{Meta-learning}
Meta-learning, or ``learning to learn,'' is a training paradigm in which a model is optimized to rapidly adapt to new tasks using limited data \cite{F1}. It mimics the human ability to transfer prior experience to new situations by learning how to learn from diverse tasks. Unlike traditional models that train on a fixed dataset and generalize within that domain, meta-learning involves exposure to a distribution of tasks during training, enabling the model to extract transferable knowledge. This knowledge allows the model to quickly fine-tune to new tasks with limited data. The approach typically involves two components: a base-learner that learns task-specific patterns, and a meta-learner that captures cross-task learning strategies to guide fast adaptation \cite{F1,R3,R4,R5}.

\par Meta-learning has been extensively explored in image processing and computer vision to enable deep-learning models with limited data samples. However, its application in the audio domain remains comparatively limited. Meta-learning mimics the human ability to learn from previous experiences or knowledge. Meta-learning algorithms are further divided into metric-based, model-based, and optimization-based approaches. References \cite{F2, F4, F5, F6, F7, F8, F11} primarily focus on metric-based few-shot learning, especially using prototypical networks. For instance, \cite{F4, F11} applies prototypical networks to drum transcription, treating each instrument as a binary classification task. Similarly, \cite{F5,F6,F7} handle individual audio event clips as binary classification tasks. In \cite{F8}, multi-class audio segments are framed as binary classification, with the target class as positive and others as negative. Furthermore, \cite{F12} applies the MAML algorithm to melody extraction.

\subsection{$T\bar{a}la$ Identification (Music Sequence Matching)}
\label{tal in HM}
In Hindustani music, a $t\bar{a}la$ defines the rhythmic structure of a composition. It comprises a fixed time cycle divided into equal time units called $m\bar{a}tras$, grouped into sections ($vibh\bar{a}gs$), forming a complete cycle known as an $avart$. The starting point of this cycle is referred to as $s\bar{a}m$, and each $t\bar{a}la$ features a characteristic rhythmic pattern known as the $thek\bar{a}$. Comprehensive explorations of $t\bar{a}la$ in Hindustani music can be found in \cite{G4, G5}. Despite fundamental differences between Hindustani and Western music, the term equivalence aids clarity. Accordingly, we adopt `beat' for $m\bar{a}tra$ and `stroke' for $b\bar{o}l$ to facilitate cross-cultural rhythmic analysis.

The tabla, consisting of two drums: the left-hand $bayan$ and the right-hand $dayan$, is the primary percussion instrument used to render the $thek\bar{a}$. A wide range of $t\bar{a}las$ exists in Hindustani music, with $T\bar{i}nt\bar{a}l$, $Ek\bar{a}l$, $Jhapt\bar{a}l$, and $Rupak\ t\bar{a}l$ being the most popular and the focus of this study. Six major $ghar\bar{a}n\bar{a}s$: Delhi, Ajrada, Lucknow, Banaras, Farrukhabad, and Punjab, define distinctive tabla styles. These styles vary in stroke articulation, dynamics, ornamentation, and improvisational phrasing, though the structure and number of strokes in the $thek\bar{a}$ typically remain consistent.

\par Accurate $t\bar{a}la$ identification is essential for tasks such as rhythm analysis, segmentation, and transcription. String-matching techniques have been extensively explored in the literature for music sequence comparison. In \cite{T1}, Longest Common Subsequence (LCS) and Rough LCS (RLCS) are used for symbolic matching. A modified RLCS approach is presented in \cite{T2} for motif detection in Carnatic $al\bar{a}panas$. In the context of tabla, \cite{S8}employ a modified RLCS approach for matching specific stroke sequences in transcribed performances. In \cite{T5}, $R\bar{a}ga$ verification is performed using Longest Common Segment Set (LCSS). While these techniques preserve element order, they do not enforce continuity, limiting their effectiveness for $t\bar{a}la$ patterns where ordered and contiguous strokes are crucial. Alternatively, \cite{T7} propose beat amplitude pattern matching, assuming a correlation between strokes and amplitude. However, performance variability in dynamics often renders these methods unreliable in real-world scenarios.

Consequently, there remains a need for matching strategies that retain both continuity and order in stroke sequences while accommodating expressive variations in timing and dynamics.

\begin{figure}[t!]
\centering
\includegraphics[scale=0.9]{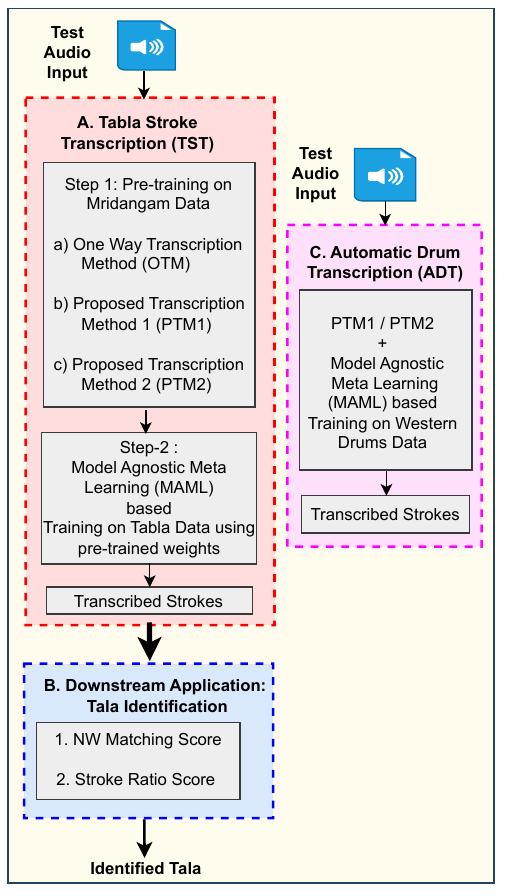}
\caption{System overview: \textbf{(A)} TST: Pre-trained on mridangam data using OTM, PTM1, PTM2, and adapted via MAML. \textbf{(B)} $T\bar{a}la$ identification: Identifies $t\bar{a}la$ using transcribed strokes from TST and NW Matching/Stroke Ratio scores. \textbf{(C)} ADT: Applies the same meta-learning strategies to Western drum audio.}
\label{proposed_system}
\end{figure}

\section{Proposed Method}\label{s_proposed_system}
This section describes our meta-learning-based system for low-resource Tabla Stroke Transcription (TST) and Automatic Drum Transcription (ADT). Fig.\ref {proposed_system} illustrates the system overview, highlighting MAML's role in enabling few-shot adaptability. The system consists of three components: A) A CRNN-based TST model trained with MAML that transcribes audio into stroke sequences. B) A downstream $t\bar{a}la$ identification module that infers rhythmic structure from these sequences. C) Validation of the framework on a Western drums dataset for the ADT task, demonstrating adaptability across diverse drum data. We introduce two novel scoring methods for $t\bar{a}la$ identification: Needleman–Wunsch (NW) Matching Score and Stroke Ratio Score. Each component is detailed in the following subsections.

\subsection{Tabla Stroke Transcription (TST)}
This section presents our solution for TST with limited labelled data. We train and test deep learning models using three methods to identify the optimal transcription approach. Subsequently, we leverage the pre-trained model weights from the best method for MAML for TST.

\subsubsection{Step1: Pre-training} \label{Step1: Training deep learning models on the Synthetic Mridangam Stroke Dataset}
In this context, we implement three distinct transcription methods, as explained below.
\paragraph{One-way Transcription Method (OTM) \cite{S9,S10}} \label{OTM}

We adopt the method of Rohit et al. \cite{S9, S10}, expanding stroke classes from four to ten, with a separate CNN model for each class. Let $(X_i, Y_i)$ be a sample from a labelled dataset of $N$ strokes. The input $X_i \in \mathbb{R}^{3 \times F \times T}$ is a 3-channel chunk of 15 time frames $(T)$ from log-Mel spectrograms computed using window sizes of 23ms, 46ms, and 93ms (10ms hop), stacked to combine multiple time-frequency resolutions for capturing both transient and tonal features. The model $f_{\theta}$ predicts the onset at the center frame by leveraging contextual information from surrounding frames, with $Y_i \in {\{0,1\}} $ indicating the presence or absence of a stroke onset.

\paragraph{Proposed Transcription Method - 1  (PTM1)}\label{PTM1}
This work proposes a new mridangam and tabla stroke transcription method inspired by Sound Event Detection (SED) tasks \cite{G10}. The proposed method utilizes a CRNN for stroke transcription. The labeled dataset of $N$ samples is denoted as $D(X_i,Y_i)$, where $i=0,1,…,N-1$. Here, $X_i \in \mathbb{R}^{F \times T}$ is the input feature matrix (standardized log Mel-spectrogram), containing $F$ acoustic features extracted from each frame in each audio recording with $T$ time frames. The labels are one-hot encoded as the target output matrix $Y_i\in \left \{ 0,1\right \}^{C \times T}$, where $C$ represents the number of predefined stroke classes.
\par Let $f_{[\theta_1,\theta_2,\theta_3]}$ be the multi-class classification model that maps the given input data $X$ to the target outputs $Y$. Here, $\theta_1$, $\theta_2$, and $\theta_3$ are the parameters of convolutional, recurrent, and classifier layers, respectively. For a given feature vector $x$ for a single time frame, $f_{[\theta_1,\theta_2,\theta_3]}$ outputs the stroke label with the highest soft-max probability. 
During the training phase, the model parameters $\theta_1$,$\theta_2$, and $\theta_3$ are randomly initialized and updated using the stochastic gradient descent algorithm as:
\begin{equation}\label{gd_algo}
    [\theta_1,\theta_2,\theta_3] \leftarrow [\theta_1,\theta_2,\theta_3] - \alpha  \nabla_{[\theta_1,\theta_2,\theta_3]} L_{wCE} \left ( f_{[\theta_1,\theta_2,\theta_3]}\right )
\end{equation}
where,  $\alpha \in \mathbb{R}^+$ is the learning rate, and $L_{wCE}$ is weighted categorical cross-entropy loss to handle the class imbalance defined as:
\begin{equation}\label{wtd ce loss}
    L_{wCE}=-\sum_{c=1}^{C} w_c y_{ic} \ \mathrm{log}(\hat{y_{ic}})
\end{equation}
where, $C$ is the number of classes, $\hat{y_i}$ is the predicted output, and $y_i$ is the ground truth for $i^{th}$ time frame for class $c$, i.e., 0 or 1 and $ w_c \in  \mathbb{R}^{+}$ is inversely proportional to the total number of time frames in training data corresponding to each class $c$.

After obtaining the frame-wise class predictions from the model, a post-processing step is applied to refine the output and extract onset times. This involves two stages. First, label smoothing is performed to correct short-term inconsistencies: for any frame $t \in [1, T-2]$, if the predicted class at $t$ differs from those at $t-1$ and $t+1$, but the labels at $t-1$ and $t+1$ are identical, then the label at $t$ is updated to match the surrounding frames, i.e., $Y_c[t] \leftarrow Y_c[t-1]$. Here, $T$ denotes the total number of time frames. Second, onset frames are identified as the time points where the predicted class changes between consecutive frames, i.e., $Y_c[t] \ne Y_c[t-1]$. These frame indices are then converted to onset times using the known frame rate, providing precise temporal localization of stroke events.

\par

\paragraph{Proposed Transcription Method - 2 (PTM2)}\label{PTM2}
This method is similar to the method discussed in Section \ref {PTM1}, utilizing the same CRNN model architecture and training procedure. However, it adopts a different target vector methodology, similar to that in \cite {D3, D5}. The targets are represented as one-hot encoded vectors, with $Y_i\in \ {\{0,1\}} ^ {C \times T} $, where $C$ denotes the number of predefined stroke classes. In this method, only the onset frame of a particular class is assigned for that class, while all non-onset frames are assigned to the `No-stroke' class. During the training phase, the model parameters are randomly initialized and updated using the stochastic gradient descent algorithm, as described by Equation (\ref{gd_algo}). A weighted cross-entropy loss function is employed as indicated by Equation (\ref{wtd ce loss}).

\subsubsection{Step 2: Model Agnostic Meta-learning (MAML)} \label{MAML explain}

The complete MAML process can be divided into meta-training and meta-testing phases. The model is trained on various tasks during meta-training to adapt to new tasks effectively with only a few samples. Additionally, the model is evaluated on different distribution/domain datasets along with different or same tasks in the meta-testing phase to assess its adaptability to unseen, diverse scenarios.

\paragraph{Meta-training} \label{meta_training}
Let, a task $T$ is sampled from a distribution $p(T)$ within the meta-training dataset $D_{train}$. Each task $T_i \sim p(T)$ is divided into a support set $T_i^s$ consisting of $s$ samples and a query set $T_i^q$  containing $q$ samples. The pre-trained base model $f_{[\theta_1, \theta_2, \theta_3]}$, with parameters $f_{[\theta_1, \theta_2, \theta_3]}$  explained in Section \ref{PTM1}, serves as the base-learner model. 
During meta-learning, we freeze the convolutional layers (pre-trained on mridangam) and update the recurrent and classifier layers. The convolutional layers capture low-level features shared between mridangam and tabla, preserving transferable features and reducing overfitting on low-resource tabla data. Thus, $\theta_1$ remains the same as in the pre-trained model while $\theta_2$ and $\theta_3$ become trainable parameters. We denote the combined parameters $\theta_2$ and $\theta_3$ as $\phi$, making $\phi$ the meta parameters of the base-learner model. The updated parameters of the base learner after $N$ steps for task $T_i$ are given by,

\begin{equation} \label{ILO_eq}
    \phi_N^i = \phi_{N-1}^{i} - \alpha  \nabla_{\phi} L_{T_i^s} \left (  f_{[\theta_1,\phi_{N-1}^i]}\right )
\end{equation}
where, $\alpha$ represents the base model's learning rate, and $L_{T_i^s} (f_ {[\theta_1, \phi_{N-1} ^i]}) $ denotes the loss computed on the support set of task $T_i$ after $(N-1)$ update steps, as given in Equation (\ref{wtd ce loss}). This process, known as inner loop optimization (ILO), involves updating the model $f_{[\theta_1, \phi]}$  based on the support set. Once base learning is complete, the model $f_{[\theta_1, \phi]} $ becomes  $f_{[\theta_1,\phi_N^i]} $. 
Subsequently, using these updated parameters $f_{[\theta_1, \phi_N^i]}$, the loss $L_{T_i^q}(\phi_N^i, T_i^q)$ is computed on the query set $T_i^q$. The meta-parameters $\phi$ are then updated using this loss. The process of updating meta-parameters over the batch of tasks is called outer loop optimization (OLO) and is given by,
\begin{equation} \label{OLO_eq}
    \phi \leftarrow \phi - \beta \nabla_{\phi}\sum_{T_i}L_{T_i^q} \left ( f_{[\theta_1,{\phi}_N^i]} \right )
\end{equation}
where $\beta$ is a meta-learning rate and $L_{T_i^q}$ is loss on the query set $T_i^q$ for task $T_i$  calculated by equation (\ref {wtd ce loss}). The entire ILO and OLO process (two-stage optimization) is repeated for all tasks $T_i$ in the $D_{train}$ dataset for $E$ epochs. The complete training process of MAML is outlined in Algorithm \ref{MAML_algo}.

\begin{algorithm}[!ht]
\caption{MAML for stroke detection}\label{MAML_algo}
    \begin{algorithmic}[1]
    \REQUIRE Pre-trained base model parameters $[\theta_1, \phi]$; frozen $\theta_1$
    \REQUIRE $\alpha$, $\beta$: learning rates

    \STATE \textbf{for} E number of epochs \textbf{do}
        \STATE \hspace{0.2cm}\textbf{for} all tasks $i$ in dataset $D_{train}$ \textbf{do}
            \STATE \hspace{0.4cm} Initialize $\phi^i = \phi$
        
                \STATE \hspace{0.4cm} Sample a batch of Log Mel spectrograms as $T_i \sim p(T)$

                \STATE \hspace{0.4cm} Select $s$ samples as a support set $T_i^s$ and  \\
                \hspace{0.4cm} $q$ samples as a query set $T_i^q$
                \STATE \hspace{0.4cm}   Update base-learner parameters $\phi_N^i$ using support set \\
                \hspace{0.4cm} $T_i^s$ by ILO ($N$ update steps) given by equation (\ref{ILO_eq})

        \STATE \hspace{0.4cm} Update $\phi$ using query set $T_i^q$ by OLO (1 update step)\\ \hspace{0.4cm} given by equation (\ref{OLO_eq})
        \STATE \hspace{0.2cm} \textbf{end for}
    \STATE \textbf{end for}
\STATE Obtained updated parameters $\phi$
    \end{algorithmic}
\end{algorithm}

\paragraph{Meta-testing}
We test the trained model $f_{[\theta_1,\phi]}$ in this stage. The updated model parameters $\phi$ from the meta-training phase now serve as good initialization parameters for adapting to new stroke classes with few samples. Given a new task $T_j$ with new unseen stroke classes from the test data $D_{test}$ consisting of a support set $T_j^s$ with $s$ samples and a query set $T_j^q$ with $q$ samples. The model $f_{[\theta_1,\phi^j]}$ is initialized with $\phi^j = \phi$ and trained on support set $T_j^s$ using equation (\ref{ILO_eq}). After $N$ update steps, the updated parameters become $\phi_N^j$. This ILO process is repeated for $E_1$ iterations. The model's performance with final updated parameters $\phi_N^j$ is then evaluated on the query set $T_j^q$.

\subsection{$T\bar{a}la$ identification}\label{tal_identify}
$T\bar{a}las$ are primarily identified by their $th\bar{e}k\bar{a}$, which maintains a fixed stroke count as discussed in Section \ref{tal in HM}. We utilize these sequence and ratio properties for $t\bar{a}la$ identification, defining two $t\bar{a}la$ identification methods (scores): `NW Matching Score' and `Stroke Ratio Score,' based on the Needleman-Wunsch algorithm \cite{T8} and Cosine Similarity, respectively.

\begin{algorithm}[!h]
\caption{Algorithm for NW Matching Score}\label{alg2}
\begin{algorithmic}[1]

\STATE Let $ X_{ref} = \left \langle x_0, x_1,...,x_{m-1} \right \rangle$ be  a reference sequence and $ Y = \left \langle y_0, y_1,...,y_{m-1} \right \rangle$ be a test sequence frame of $m$ strokes. $S\in \mathbb {Z}^{(m+1)X (m+1)}$ be  NW Matching Score matrix, where $S[i][j]$ is the matching score of the first $i$ strokes of $X_{ref}$ with the first $j$ strokes of $Y$.
\STATE Initialization:\\
$S[0][0] = 0$, \\
$S[i][0] = S[i-1][0] + gap\  penalty$ , $\forall i = 1,...,m$, \\
$S[0][j]= S[0][j-1] + gap\  penalty$ , $\forall j = 1,...,m$, \\
where, $gap\ penalty$ is `$-2$'.
\STATE  NW Matching Score matrix, $\forall \ i,j = 1,...,m$,
\begin{equation*}
    S[i][j] = max 
    \begin{cases}
 & \text{$S[i-1][j-1] + match\ score$} \\ 
 & \text{$S[i-1][j] + gap\  penalty $} \\ 
 & \text{$S[i][j-1] + gap\  penalty $} \\
\end{cases}
\end{equation*}
 where, the $match\  score$ is `$1$' if stroke $x_{i-1}$ and $y_{j-1}$ are same and `$-1$' if both are different  
 \STATE The optimum score is the sum of the scores along the optimal path obtained by backtracking from $S[m][m]$ to $S[0][0]$.
\end{algorithmic}
\label{conf1}
\end{algorithm}

\begin{figure}[!h]
\centering
\includegraphics[scale=0.47]{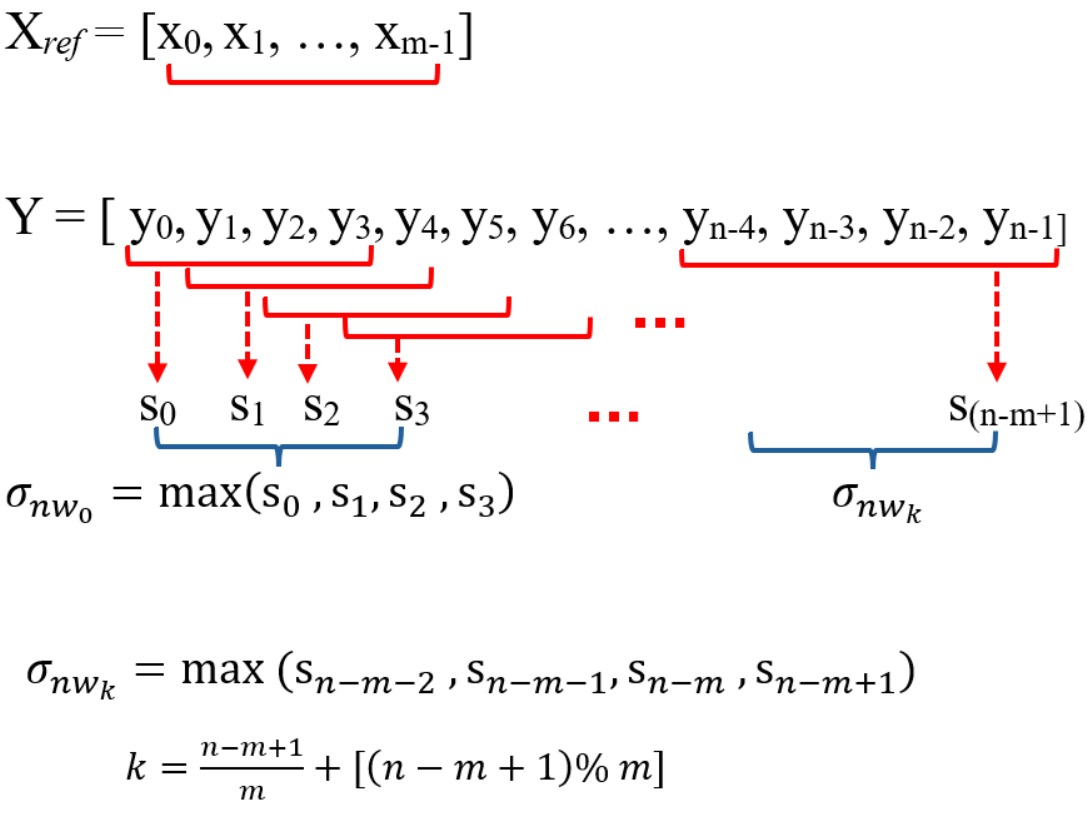}
\caption{The  NW Matching Score compares a transcripted sequence of $n$ strokes with a reference sequence of $m$ strokes.}
\label{stroke_sequence_score_nw}
\end{figure}

\subsubsection{NW Matching Score}
We obtain a stroke sequence $Y=\left\langle y_0, y_1,...,y_{n-1}\right\rangle$ of $n$ strokes from the tabla stroke transcription module. We compare it with a reference $t\bar{a}la$ sequence  $X_{ref} = \left \langle x_0, x_1,..., x_{m-1} \right \rangle$ of $m$ strokes. Since the test audio can be taken from any part of the complete audio, it doesn’t need to start from the first stroke of $X_{ref}$ (i.e., $s\bar{a}m$). We use a frame-shifting approach, sliding an $m$-stroke frame over $n$ strokes, as shown in Fig. \ref{stroke_sequence_score_nw}. The score calculation for the frame of $m$ stroke sequence is shown in Algorithm \ref{alg2}. We compute the maximum score of every $m$ shift. The final $t\bar{a}la$ matching score $\sigma_{nw}$ averages all maximum scores over $n$ strokes, given by
\begin{equation}\label{nw_eq}
    \sigma_{nw} = \frac{\sigma_{nw_{0}}+\sigma_{nw_{1}}+...+\sigma_{nw_{k}}}{k}
\end{equation}
where $k$ are steps to cover transcription sequence. The computational details of $\sigma_{nw_{0}},\sigma_{nw_{1}},...,\sigma_{nw_{k}}$ are shown in Fig. \ref{stroke_sequence_score_nw}.
This approach addresses potential missing strokes in specific frames of the test sample.

\subsubsection{Stroke Ratio Score}
We obtain the stroke count from the tabla stroke transcription module and calculate the stroke ratio for the given audio file. We analyze the beat proportions in four $t\bar{a}las$ as follows: $T\bar{i}nt\bar{a}la$ with $[Dha, Dhin, Tin, Ta] = [3,3,1,1]$,  $\bar{E}kt\bar{a}la$ with $[Dhin,Tun,Na,Kat,Ta,Dhage,Tirkita] = [3,1,2,1,1,2,2]$,  $Jhapt\bar{a}la$ with $[Dhi,Na,Ti] = [5,4,1]$, and $R\bar{u}pak \ t\bar{a}la$ with $[Tin, Na, Dhi]= [2,3,2]$. To identify the best match, we compute the cosine similarity between the test audio stroke ratio and the reference ratios for each $t\bar{a}la$. The Stroke Ratio Score is calculated as follows,
\begin{equation}
    Stroke\  Ratio\ Score = \frac{R.T}{\left \| R \right \|\left \| T \right \|}
\end{equation}
where, $R$ is the reference stroke ratio vector, and $T$ is the test stroke ratio vector.

\subsection{Automatic Drum Transcription (ADT)}
This section presents our ADT solution under limited labelled data conditions. We adopt MAML for this task, following a process similar to that described in Section~\ref{MAML explain}. A CRNN model serves as the base learner for multi-label classification. The input feature matrix is a standardized log Mel-spectrogram, consistent with those used in PTM1 and PTM2, and the target vector is constructed following the approach in \cite{D3, D5}. Unlike in TST, where pre-trained weights are used, no relevant pre-trained model exists for Western drum kits in ADT. Therefore, the entire model—including convolutional, recurrent, and linear layers—is trained from scratch, with all parameters $\theta_1, \theta_2, \theta_3$ kept trainable. where convolutional layer parameters are frozen during meta-training and meta-testing, all parameters $\theta_1, \theta_2, \theta_3$ are trainable in ADT. These parameters are optimized using a weighted binary cross-entropy loss and updated via stochastic gradient descent.

\section{Datasets}\label{Datasets}
We use diverse datasets for TST and ADT, including Only Drum/Tabla (DTD), Drums with Percussion (DTP), and Complete Concert Mixes (DTM), comprising drums/tabla alongside other percussion, melodic instruments, and vocals. Tables~\ref{data_TST} and \ref{data_ADT} summarize the datasets, detailing notations, durations, number of tracks/compositions, stroke classes, and data usage for meta-training/testing.

\par While some stroke or instrument classes (e.g., $Dha$, $Dhi$, snare, hi-hat) appear across datasets, they are treated as distinct during meta-learning due to differences in recording conditions, background instrumentation, and data distribution. Even with identical labels, contextual variations (e.g., presence of accompanying instruments or vocals) require treating them separately to ensure robust domain adaptation. Audio samples for meta-training and testing were randomly selected as per durations listed in Tables~\ref{data_TST} and~\ref{data_ADT}.

\begin{table*}[b]
\centering
\caption{Meta-Learning Dataset Configuration for Tabla Stroke Transcription (TST)}
\label{data_TST}
\setlength{\tabcolsep}{6pt}
\begin{tabular}{@{}lcccl@{}}
\toprule
\multicolumn{1}{c}{{\textbf{Dataset}}} & {\textbf{Notation}} & {\textbf{Total Dur. (hr)}} & {\textbf{Stroke Classes Used}} & \multicolumn{1}{c}{{\textbf{Meta Train-Test Data Usage \& Evaluation}}} \\ 
\midrule
{\textbf{\begin{tabular}[c]{@{}l@{}}Synthetic Mridangam \\ Stroke Dataset\end{tabular}}} & {$D_M$} & {\begin{tabular}[c]{@{}l@{}}\hspace{0.35cm} 2.0\\ (720 excerpts)\end{tabular}} & {\begin{tabular}[c]{@{}c@{}}$Thi, Ta, Num, Tha, Dhin, Cha,$ \\ $Bheem, Thom, Tham, Dheem$\end{tabular}} &  \multicolumn{1}{c}{{Used in Pre-training stage}} \vspace{0.2cm} \\

{\textbf{\begin{tabular}[c]{@{}l@{}}Tabla Solo \\ Dataset \cite{S8}\end{tabular}}} & {$D_{T1}$} & {\begin{tabular}[c]{@{}l@{}}\hspace{0.3cm} 0.29\\ (38 composi- \\ \hspace{0.3cm} -tions)\end{tabular}} & {\begin{tabular}[c]{@{}c@{}}$Da, Ki, Ge, Kda, Tit, Dha, Dhe,$ \\ $Dhi, Re, Tin, Ta, Na, Din, Dhet,$ \\ $Dhin$\end{tabular}} & {\begin{tabular}[c]{@{}l@{}}\textbf{Meta-Train}: 12 min (0.2 hr) and 10 classes \\ ($Da, Ki, Ge, Kda, Tit, Dha, Dhe, Dhi, Re, Tin$)\\ \textbf{Meta-Test-1}: 5 min (0.08 hr) adaptation on 5 new classes \\ ($Ta, Na, Din, Dhet, Dhin$); Evaluation on 0.21 hr \vspace{0.2cm}\end{tabular}} \\

{\textbf{\begin{tabular}[c]{@{}l@{}}4way-tabla-Dataset\\ ismir21 \cite{S9}\end{tabular}}} & {$D_{T2}$} & {\begin{tabular}[c]{@{}l@{}}\hspace{0.3cm} 1.6\\ (38 tracks)\end{tabular}} & {B, D, RB, RT} & {\begin{tabular}[c]{@{}l@{}}\textbf{Meta-Test-2}: 10 min (0.17 hr) adaptation on new classes; \\ Evaluation on a 20 min (0.33 hr) separate test set by \cite{S9} \vspace{0.2cm}\end{tabular}}  \\

{\textbf{\begin{tabular}[c]{@{}l@{}}Hindustani Music\\ Rhythm Dataset \cite{G4}\end{tabular}}} & {$D_{T3}$} & {\begin{tabular}[c]{@{}l@{}}\hspace{0.5cm} 5.03\\ (151 excerpts)\end{tabular}} & {\begin{tabular}[c]{@{}c@{}}$Dha, Dhim, Tin, Na, Tun, Kat,$ \\ $Ta, Dhage, Tirikita, Dhi, Ti$\end{tabular}} & {\begin{tabular}[c]{@{}l@{}}\textbf{Meta-Test-3}: 10 min (0.17 hr) adaptation on new classes; \\ Evaluation on 4.86 hr \end{tabular}} \\
\bottomrule
\end{tabular}
\end{table*}

\begin{table*}[b]
\centering
\caption{Meta-Learning Dataset Configuration for Automatic Drum Transcription (ADT)}
\label{data_ADT}
\setlength{\tabcolsep}{3.5pt}
\resizebox{\textwidth}{!}{%
\begin{tabular}{@{}lccccl@{}}
\toprule
\multicolumn{1}{c}{{\textbf{Dataset}}} &
{\textbf{Notation}} & 
{\textbf{ADT Task Type}} & 
{\textbf{Total Dur. (hr)}} & 
{\textbf{Stroke Classes Used}} & 
{\textbf{Meta Train-Test Data Usage \& Evaluation}} \\ 
\midrule

{\textbf{ADTOF \cite{D16}}} & 
{$D_{D4}$} & 
{DTM} & 
{\begin{tabular}[c]{@{}l@{}} \hspace{0.5cm}114\\ (1739 tracks) \end{tabular}} & 
{BD, SD, TT, HH, CY+RD} & 
{\textbf{Meta-Train}: on full dataset (all classes seen)} \\

{\textbf{\begin{tabular}[c]{@{}l@{}}IDMT-SMT-\\ Drums \cite{D14}\end{tabular}}} & 
{$D_{D1}$} & 
{DTD} & 
{\begin{tabular}[c]{@{}l@{}}\hspace{0.35cm} 2.04\\ (104 tracks) \end{tabular}} & 
{KD, SD, HH} & 
{\begin{tabular}[c]{@{}l@{}}\textbf{Meta-Test-1}: 5-min (0.083hr) adaptation\\ on new classes; Evaluation on 1.96 hr \vspace{0.2cm}\end{tabular}} \\

{\textbf{\begin{tabular}[c]{@{}l@{}}ENST-Drums\\ minus-one \cite{D13}\end{tabular}}} & 
{$D_{D2}$} & 
{DTP} & 
{\begin{tabular}[c]{@{}l@{}}\hspace{0.3cm} 1.01\\ (64 tracks) \end{tabular}} & 
{\begin{tabular}[c]{@{}c@{}}BD, SD, CL, HH,\\ BE, TT, RD, CY\end{tabular}} & 
{\begin{tabular}[c]{@{}l@{}}\textbf{Meta-Test-2}: 5-min (0.083 hr) adaptation\\ on new classes; Evaluation on 0.927 hr \vspace{0.2cm}\end{tabular}} \\

{\textbf{MDB-Drums \cite{D15}}} & 
{$D_{D3}$} & 
{DTD, DTM} & 
{\begin{tabular}[c]{@{}l@{}}\hspace{0.25cm} 0.35\\ (23 tracks) \end{tabular}} & 
{\begin{tabular}[c]{@{}c@{}}KD, SD, HH, HH-O,\\ TT, CY, RD, OT\end{tabular}} & 
{\begin{tabular}[c]{@{}l@{}}\textbf{Meta-Test-3}: 5-min (0.083 hr) adaptation\\ on new classes; Evaluation on 0.267 hr \end{tabular}} \\
\bottomrule
\end{tabular}%
}

\end{table*}

\subsection{Synthetic Mridangam Stroke Dataset $(D_M)$}\label{mrid_data} 

The mridangam is a South Indian barrel-shaped percussion instrument with treble and bass heads, offering a wide tonal range. It is traditionally carved from jackfruit wood and fitted with animal skin membranes. We curated a dataset of 720 ten-second audio files by randomly concatenating strokes from the Mridangam Stroke Dataset \cite{G1}, which contains 10 stroke classes across 6 tonic pitches, recorded at 44.1 kHz. The dataset includes 120 files per tonic, totalling 120 minutes of audio.

\section{Experimental Setup} \label{Exp_set_up}

\begin{figure}[!h]
\centering
\includegraphics[scale=0.72]{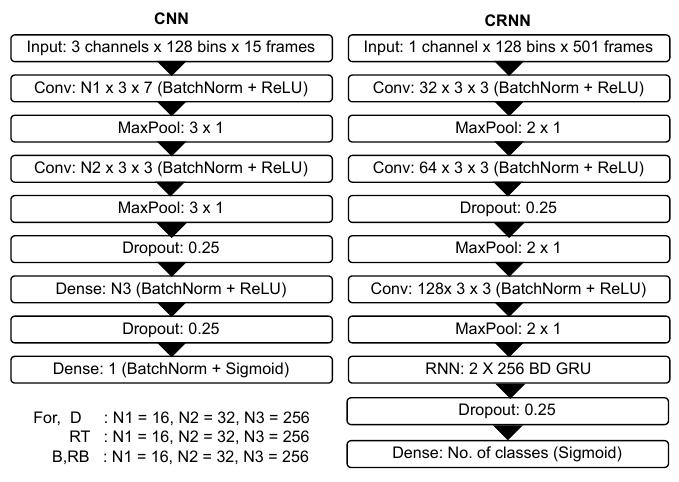}
\vspace{-0.2cm}
\caption{General CNN \cite{S10} and CRNN model architecture for mridangam and tabla stroke transcription}
\label{CNN_CRNN}
\end{figure}

The experimental setup follows the same flow as shown in Fig.\ref{proposed_system} and consists of two major components: TST and ADT. Across all experiments, we employ simple deep CNN and CRNN models for mridangam and tabla stroke transcription, as well as for the ADT task. The architectures of the CNN and CRNN models are illustrated in Fig.\ref{CNN_CRNN}.

\subsection{Tabla Stroke Transcription (TST)}
\subsubsection{Step - 1: Pre-training on $D_M$}
\paragraph{One-way Transcription Method (OTM)} \label{exp_OTM}
We employ 10 CNNs for 10 classes as discussed in \ref{mrid_data}. Akshay et al. \cite{G2} categorized these classes into damped (D), resonant-treble (RT), resonant-bass (RB), and resonant-both (B). We use the same model architectures as Rohit et al. \cite{S9} for these ten classes as follows: for $Thi$, $Ta$, $Num$, $Tha - $D; for $Dhin$, $Cha$, $Bheem - $ RT; for $Thom - $RB; and for $Tham$, $Dheem - $B. To achieve tonic-independent inference, we applied 6-fold cross-validation, splitting the dataset into six folds by tonic. We trained models on five folds and tested on one, with each fold's training data split 90:10 for training and validation. This process covered all six tonics, and we averaged the evaluation scores.

\par The CNN, trained on 3-channel log-Mel spectrogram inputs, uses the Adam optimizer with a batch size of 256 for up to 150 epochs. Early stopping with a patience of 10 epochs is applied, and the model with the lowest validation loss is retained. During inference, a fixed-threshold peak picking algorithm \cite{S13} is used to binarize the network output, identifying significant peaks and converting the continuous activations into binary decisions.

\paragraph{Proposed Transcription Method - 1 (PTM1)}\label{exp_PTM1}

The datasets used in this work vary in audio length, so we segment them into non-overlapping 5-second chunks for feature extraction, each labelled accordingly. The input to the CRNN is a standardized log Mel spectrogram, computed with a 46.4 ms window size and a 10 ms hop size, utilizing 128 Mel filter banks. Each frequency band of the Mel spectrogram is normalized to have zero mean and unit variance. This method assigns equal importance to the duration of the entire stroke rather than focusing solely on the onset frame.
 
\par
Mridangam and tabla strokes are primarily characterized by the attack, decay, and sustain phases of the ADSR model, with most strokes having a fast-decaying exponential envelope \cite{S1}. Therefore, stroke onset is predominantly defined by the energetic attack-decay phase, while the sustain phase defines how the stroke ends. The last part of the release region contributes limitedly to defining stroke onset and class. If an equal focus is given to the complete stroke duration during transcription, it can confuse the model since this phase is similar across all stroke classes. Hence, we transcribe this segment separately as `No-stroke' using a threshold set at 3\% of the stroke's maximum amplitude. Frames from onset to threshold denote the stroke class, while frames from threshold to the subsequent stroke onset are labelled `No-stroke.' We employ 6-fold cross-validation using six tonics. Training involves a batch size of 32 over 100 epochs, with a learning rate of 0.001. During inference, class-wise onsets are separated, and only ten stroke classes are considered by eliminating the `No-stroke' class for evaluation.

\paragraph{Proposed Transcription Method - 2 (PTM2)}
The CRNN model architecture and standardized log Mel spectrogram input to the CRNN are the same as explained in Section \ref{exp_PTM1}. The training targets include the `No-stroke' class, as detailed in Section \ref{PTM2}. The model is trained with a batch size of 32 for 100 epochs, using a learning rate of 0.001. During inference, the model predicts stroke classes frame-by-frame for a given test log-Mel spectrogram. The onset of each stroke class corresponds to the frame in which the model predicts it. Then, class-wise onsets are separated, and only ten stroke classes are considered by eliminating the `No-stroke' class.

\subsubsection{Step 2}
Based on the pretraining results, OTM performs poorly in complexity and time consumption, requiring a CNN for each stroke class. As a result, we will discontinue this approach and instead focus on the proposed methods. Specifically, we will employ the CRNN model for MAML-based domain adaptation for tabla strokes. Our methodology will be validated across three tabla datasets.

\paragraph{MAML}\label{exp_set MAML}

We meta-train our models on ten stroke classes from $D_{T1}$ and evaluate in three meta-testing scenarios involving $D_{T1}$, $D_{T2}$, and $D_{T3}$. The corresponding stroke class distributions and audio durations for both meta-training and testing are detailed in Table~\ref{data_TST}. All datasets come from different distributions and recording backgrounds. Additionally, the model adapted on $D_{T2}$ during meta-testing is directly evaluated on a separate test partition from $D_{T2}$ provided by Rohit et al.~\cite{S9}, allowing comparison with the baseline.

\par In the first scenario, we must partition the existing classes into two mutually exclusive sets. Due to the challenge of directly segmenting long audio based on stroke classes, we adopt an approach from Nolasco et al. \cite{F2} and Shi Bowen et al. \cite{F8}. Here, we preserve the original class label for desired strokes and classify non-target strokes as `Others'. As a result, there will be differences in the classes used between meta-training and meta-testing. The `No-stroke' class is also considered in the `PTM1' and `PTM2', discussed earlier in Sections \ref{PTM1},\ref{exp_PTM1}, and \ref{PTM2}. 
\par The MAML approach is applied to both methods with the same experimental settings, utilizing their respective input-output pairs. For task $T_i$, we randomly sample log Mel spectrograms with a batch size comprising 32 support samples and 8 query samples from the meta-training dataset. Models are trained with the Adam optimizer for 300 epochs, employing an inner loop optimization of $N = 3$. The learning rates $\alpha$ and $\beta$ are fixed at 0.001.

To evaluate the effectiveness of our proposed method against existing approaches, we conducted comparative experiments using a standard supervised training setup. Specifically, we trained and tested the same CRNN model using the PTM1 and PTM2 paradigms, with limited training data matching the meta-adaptation setup. We also evaluated a transfer learning baseline, where the model was fine-tuned and tested using the same adaptation and evaluation data. All experiments—supervised, transfer learning, and meta-learning—were performed on identical train-test splits to ensure fair comparison. The models were trained with a batch size of 32 for 100 epochs using a learning rate of 0.001.

\subsubsection{Baseline Models for TST Comparison}
To evaluate the effectiveness of our proposed method, we conducted comparative experiments using a standard supervised setup. We trained and tested the same CRNN model with PTM1 and PTM2, using limited data matching the meta-adaptation setup. We also evaluated transfer learning, where the model was fine-tuned and tested with the same data. All experiments (supervised, transfer learning, and meta-learning) used identical train-test splits for a fair comparison. The models were trained with a batch size of 32 for 100 epochs at a learning rate of 0.001.

\subsection{Automatic Drum Transcription (ADT)}
\subsubsection{MAML for ADT}\label{MAML_ADT}

Meta-training is performed using five stroke classes from $D_{D4}$, while meta-testing is conducted across three scenarios involving $D_{D1}$, $D_{D2}$, and $D_{D3}$. The specific stroke class distributions and audio durations for both meta-training and testing are summarized in Table~\ref{data_ADT}. 
For evaluation, the model trained during meta-testing is directly tested on the remaining portions of each dataset to compare its performance against the baseline approach. The MAML framework follows the same experimental setup as in the tabla stroke transcription. 

\subsubsection{Baseline Models for ADT Comparison}
For ADT baseline comparisons, we evaluated three approaches: a simple CRNN-based classifier \cite{D7}, an NMF-based method following the setup described in~\cite{R2}, and a transfer learning baseline. NMF templates were derived from $D_{D1}$ and $D_{D2}$ and adapted across all datasets. To assess the impact of source separation, we applied Demucs~\cite{R1} to separate mixed audio and extracted the drum track, on which all methods were re-evaluated. All experiments used the same set of stroke classes and train-test splits as in the MAML setup to ensure fair comparison.

\begin{table*}[t]
\begin{minipage}{0.58\textwidth}
   \centering
\begin{threeparttable}
\caption{ Step-1: Transcription Performance on $D_M$ (\% f1-score)}
\label{table_D_M}
\setlength{\tabcolsep}{2pt} 
\begin{tabular}{@{}lccccccccccc@{}}
\toprule
\textbf{Approach} & \textbf{$Bheem$} & \textbf{$Cha$}  & \textbf{$Dheem$} & \textbf{$Dhin$} & \textbf{$Num$}  & \textbf{$Ta$}   & \textbf{$Tha$}  & \textbf{$Tham$} & \textbf{$Thi$}  & \textbf{$Thom$} & \textbf{Avg.} \\ \midrule
\textbf{OTM}      & 83.2           & 47.3          & 55.6           & 51.6          & 76.4          & 64.8          & 70.0            & 73.3          & 61.1          & 67.1          & 65.04            \\
\textbf{PTM2}     & 95.4           & 86.7          & 87.3           & 78.2          & 91.6          & 85.9          & 91.4          & 83.3          & 85.3          & 86.5          & 87.16            \\
\textbf{PTM1}     & \textbf{96.6}  & \textbf{93.6} & \textbf{94.1}  & \textbf{92.5} & \textbf{94.6} & \textbf{92.2} & \textbf{95.1} & \textbf{89.9} & \textbf{93.4} & \textbf{89.7} & \textbf{93.17}   \\ \bottomrule
\end{tabular}
\begin{tablenotes}
    \item OTM: One-way Transcription Method, PTM1:Proposed Transcription Method-1, PTM2:Proposed Transcription Method-2
   \end{tablenotes}
       \end{threeparttable}%
\end{minipage}%
\hspace{2em}
\begin{minipage}{0.34\textwidth}
 \centering
 \captionsetup{justification=centering}
\caption{TST Performance on $D_{T1}$ \\\hspace{1cm}(\% f1-score)}
\label{table_D_T1}
\setlength{\tabcolsep}{2.4pt} 
\begin{tabular}{@{}lccccccc@{}}
\toprule
\textbf{Approach} & \textbf{$Other$} & \textbf{$Ta$} & \textbf{$Na$} & \textbf{$Din$} & \textbf{$Dhet$} & \textbf{$Dhin$} & \textbf{Avg.} \\ \midrule
\textbf{PTM2}    & 28.2           & 30.7         & 26.3        & 31.4         & 29.5          & 30.3          & 29.40  \\
\textbf{PTM1}     & 34.9           & 36.8         & 33.1        & 38.5        & 35.7          & 37.6          & 36.11   \\
\textbf{PTM2+TL}  & 69.1           & 48.7         & 50.4        & 51.9         & 58.1          & 53.1          & 55.22   \\
\textbf{PTM1+TL}  & 80.1           & 58.9         & 54.9        & 57.7         & 59.9          & 61.5          & 62.17   \\
\textbf{PTM2+ML}  & 82.4           & 58.8         & 69.8        & 89.3         & 91.4          & 67.5          & 76.53   \\
\textbf{PTM1+ML}  & \textbf{87.0}             & \textbf{71.2}         & \textbf{76.1}        & \textbf{89.9}         & \textbf{92.1}          & \textbf{71.6}          & \textbf{81.32}   \\ \bottomrule
\end{tabular}
\begin{tablenotes}
    \item TL: Transfer Learning, ML: Meta Learning
   \end{tablenotes}

\end{minipage}
\end{table*}

\begin{table}[h]
\centering
\caption{TST Performance on $D_{T2}$ (\% f1-score)}
\label{table_D_T2}
\setlength{\tabcolsep}{7.5pt}  
\begin{tabular}{@{}lccccc@{}}
\toprule
\textbf{Approach}                                & \textbf{B}    & \textbf{D}    & \textbf{RB}   & \textbf{RT}   & \textbf{Avg.} \\ \midrule
\textbf{Drum-pretrained  \cite{S10}}                 & 2.1           & 44.0            & 17.5          & 48.7          & 28.05            \\
\textbf{Best  set    of  D,  RT,    RB  \cite{S10}}  & 81.5          & 83.0            & 63.6          & 86.0            & 78.52            \\
\textbf{Retrained  \cite{S10}}                        & 82.7          & 83.6          & 66.9          & 86.6          & 79.95            \\
\textbf{Retrained  Rohit \cite{S9}}             & 80.1          & 83.3          & 34.1          & 84.3          & 70.40             \\
\textbf{PTM2}                                    & 36.1          & 37.2          & 23.5          & 40.7          & 34.38            \\
\textbf{PTM1}                                   & 43.8          & 46.8          & 28.6          & 47.7          & 41.73            \\
\textbf{PTM2  +  TL}                             & 57.9          & 59.0            & 35.6          & 65.6          & 54.52            \\
\textbf{PTM1  +  TL}                             & 59.4          & 72.1          & 38.6          & 71.0            & 60.28            \\
\textbf{PTM2  +  ML}                             & 81.0            & 79.6          & 63.7          & 82.9          & 76.80             \\
\textbf{PTM1  +  ML}                             & \textbf{87.3} & \textbf{85.8} & \textbf{69.4} & \textbf{89.9} & \textbf{83.10}    \\ \bottomrule
\end{tabular}
\end{table}

\begin{table*}[t]
  \begin{minipage}{0.68\textwidth}
\centering
\caption{TST Performance on $D_{T3}$ (\% f1-score)}
\label{table_D_T3}
\setlength{\tabcolsep}{3.5pt} 
 \resizebox{\columnwidth}{!}{%
\begin{tabular}{@{}lcccccccccccc@{}}
\toprule
\textbf{Approach} & \textbf{$Dha$}  & \textbf{$Dhin$} & \textbf{$Tin$} & \textbf{$Na$}   & \textbf{$Tun$} & \textbf{$Kat$}  & \textbf{$Ta$}  & \textbf{$Dhage$} & \textbf{$Tirkita$} & \textbf{$Dhi$}  & \textbf{$Ti$}  & \textbf{Avg.} \\ \midrule
\textbf{PTM2}     & 15.5          & 25.5          & 8.5           & 21.0           & 14.5          & 17.2          & 15.9          & 21.4           & 17.6              & 14.3          & 8.8           & 16.38            \\
\textbf{PTM1}     & 22.7          & 31.0            & 13.5          & 27.4          & 23.5          & 20.2          & 22.8          & 24.2           & 24.6              & 19.3          & 10.2          & 21.76            \\
\textbf{PTM2+TL}  & 19.2          & 36.9          & 9.9           & 34.8          & 16.3          & 21.3          & 26.4          & 39.2           & 27.9              & 22.8          & 23.1          & 25.25            \\
\textbf{PTM1+TL}  & 24.3          & 45.8          & 24.0            & 50.7          & 61.5          & 26.8          & 48.7          & 47.6           & 45.2              & 33.5          & 26.1          & 39.47            \\
\textbf{PTM2+ML}  & 39.4          & 61.2          & 31.3          & 45.5          & 49.0            & 65.2          & 65.6          & 64.1           & 61.2              & 46.9          & 38.1          & 51.59            \\
\textbf{PTM1+ML}  & \textbf{45.5} & \textbf{66.9} & \textbf{38}   & \textbf{55.2} & \textbf{87.8} & \textbf{71.6} & \textbf{81.4} & \textbf{78.2}  & \textbf{78.7}     & \textbf{49.5} & \textbf{40.5} & \textbf{63.02}   \\ \bottomrule
\end{tabular}

    }
  \end{minipage}
  \hspace{1em}
  \begin{minipage}{0.3\textwidth}
    \centering
    \captionsetup{justification=centering}
\caption{$T\bar{a}la$ Identification Performance on $D_{T3}$}
    \centering
\label{tal_seq}
    \centering
    \setlength{\tabcolsep}{3.5pt} 
    \resizebox{\columnwidth}{!}{%
\begin{tabular}{@{}lcc@{}}
\toprule
\textbf{Method }            & \textbf{\begin{tabular}[c]{@{}l@{}}Accuracy\\ \hspace{0.22cm} (\%)\end{tabular}} & \textbf{\begin{tabular}[c]{@{}l@{}}Time \\ \hspace{0.1cm}(ms)\end{tabular}}    \\ \midrule
\textbf{RLCS$_{0}$ {[}9{]} }   & 33.6          & 64.2          \\
\textbf{NW Matching Score } & \textbf{48.9} & 62.77         \\
\textbf{Stroke Ratio Score }& 43.1          & \textbf{1.41} \\ \bottomrule
\end{tabular}
    }
  \end{minipage}
\end{table*}

\section{Results and Discussions}\label{results}

Performance is evaluated using the f1-score with a 50 ms collar for detecting onset positions. Scores are computed for each stroke type on individual tracks and then averaged across the dataset using the $mir\_eval$ Python library for both TST and ADT.

\subsection{Tabla stroke transcription (TST)}

Table~\ref{table_D_M} presents the average cross-validation results of pre-training on $D_M$ using the methods from Section \ref{Step1: Training deep learning models on the Synthetic Mridangam Stroke Dataset}. Predictions are obtained by thresholding local peaks in network activations, with thresholds tuned on the validation set. The proposed methods outperform OTM, which captures only frequency-domain features. In contrast, the CRNN leverages both spectral and temporal information, modelling the ADS characteristics of strokes. PTM1, which uses all frames within a stroke, outperforms PTM2, which relies solely on onset frames—likely affected by annotation inconsistencies. Leveraging temporal context, PTM1 offers improved robustness. Additionally, CRNN models perform well with single-channel input, indicating that OTM’s 3-channel configuration is not essential.

Tables~\ref{table_D_T1}–\ref{table_D_T3} show results for PTM1 and PTM2 under meta-learning across test sets with varying data distributions. PTM1 consistently outperforms PTM2 and other baselines, demonstrating the advantage of MAML’s rapid task adaptation over fixed representations. Table~\ref{table_D_T2} highlights PTM1 outperforming four OTM settings, emphasizing CNN's limitations in modeling temporal dependencies. Performance drop on $D_{T3}$ (Table~\ref{table_D_T3}) is due to stroke overlap with vocals and instruments, hindering onset detection.

\par The comparison with NMF-based and source separation methods is excluded for TST due to several limitations. Unlike ADT, TST lacks isolated instrument recordings, and using templates from the test dataset would create an unfair comparison. Additionally, no publicly available source separation model is designed for Indian classical music, and applying existing models like Demucs \cite{R1} leads to artifacts and domain mismatch, making them unsuitable for accurate stroke-level transcription.
\subsection{$T\bar{a}la$ identification}
After TST, $t\bar{a}la$ identification involves generating $t\bar{a}la$ identification scores. The average processing times for this task from transcribed tabla stroke sequences are 1.41 ms for cosine similarity scoring and 62.77 ms for sequence matching score when analyzing 2 minutes of test audio across 4 $t\bar{a}las$. These computations were performed on a system equipped with 16GB RAM, an Intel i7 processor, and a 6GB GPU.

\subsection{Automatic drum transcription (ADT)}

Tables~\ref{table_DD1}–\ref{table_DD3_2} present the performance of the proposed MAML-based approach across all ADT task types—DTD, DTP, and DTM—compared with baselines including the state-of-the-art CRNN model \cite{D7},  an NMF-based method \cite{R2}, and transfer learning. For the DTM dataset, source separation \cite{R1} is applied, and results are reported accordingly. Table~\ref{table_mdb_all} further compares MAML on $D_{D3}$ (DTM) against existing methods, using a 30ms collar for fair evaluation.

Among the baselines, the NMF-based method performs better than the supervised CRNN in low-data scenarios, showing greater robustness under limited supervision. However, its performance degrades on the DTM dataset due to interference from other instruments and vocals, which hampers reliable template construction in the absence of isolated tracks.

Overall, the MAML-based approach consistently outperforms all baselines across ADT tasks, demonstrating strong generalization and robustness in low-resource and real-world conditions.

\begin{table}[h]
\centering
\setlength{\tabcolsep}{8pt}  
\captionof{table}{ADT Performance on DTD - $D_{D1}$ (\% f1-score)}
\label{table_DD1}
\begin{tabular}{@{}lcccc@{}}
\toprule
\textbf{Approach}  & \textbf{KD}   & \textbf{SD}   & \textbf{HH}   & \textbf{Avg.} \\ \midrule
\textbf{SOTA-CRNN \cite{D7}} & 42.1          & 65.3          & 58.8          & 55.4             \\
\textbf{PF-NMF \cite{R2}}   & 68.4          & 72.5          & 73.0            & 71.3             \\
\textbf{CRNN + TL} & 54.3          & 68.7          & 63.9          & 62.3             \\
\textbf{CRNN + ML} & \textbf{73.9} & \textbf{84.2} & \textbf{82.1} & \textbf{80.1}    \\ \bottomrule
\end{tabular}   
\end{table}

\begin{table}[h]
\centering
\label{table_DD2}
\setlength{\tabcolsep}{3pt}  
\captionof{table}{ADT Performance on DTP - $D_{D2}$ (\% f1-score)}
\begin{tabular}{@{}lccccccccc@{}}
\toprule
\textbf{Approach}  & \textbf{BD}   & \textbf{SD}   & \textbf{CL}   & \textbf{HH}   & \textbf{BE}   & \textbf{TT}   & \textbf{RD} & \textbf{CY}   & \textbf{Avg.} \\ \midrule
\textbf{SOTA-CRNN \cite{D7}} & 54.4          & 53.1          & 41.5          & 56.5          & 34.7          & 42.6          & 47.5        & 39.1          & 46.2             \\
\textbf{PF-NMF \cite{R2}}    & 77.4          & 66.9         & 69.8          & 74.5         & 63.2          & 71.8        & 69.1       & 68.3        & 70.1             \\
\textbf{CRNN + TL} & 68.7          & 65.3          & 52.1          & 64.4          & 48.3          & 61.5          & 58.4        & 53.5          & 59.0             \\
\textbf{CRNN + ML} & \textbf{83.5} & \textbf{82.8} & \textbf{72.1} & \textbf{85.3} & \textbf{68.5} & \textbf{77.2} & \textbf{78.0} & \textbf{73.9} & \textbf{77.6}    \\ \bottomrule
\end{tabular}
    
\end{table}

\begin{table}[h]
\centering
\caption{ADT Performance on DTD - $D_{D3}$ (\% f1-score)}
  \label{table_DD3_1}
\setlength{\tabcolsep}{3pt}  
    \begin{tabular}{@{}lccccccccc@{}}
    \toprule
    \textbf{Approach}  & \textbf{KD}   & \textbf{SD}   & \textbf{HH}   & \textbf{HH-O} & \textbf{TT}   & \textbf{CY}   & \textbf{RD}   & \textbf{OT}   & \textbf{Avg.} \\ \midrule
    \textbf{SOTA-CRNN \cite{D7}} & 43.0 & 44.5 & 37.2 & 31.8 & 26.3 & 37.7 & 31.3 & 23.5 & 34.4 \\
    \textbf{PF-NMF \cite{R2}} & 59.2 & 65.1 & 71.8 & 63.3 & 68.2 & 66.1 & 63.5 & 32.2 & 61.2 \\
    \textbf{CRNN + TL} & 53.0 & 68.3 & 62.0 & 58.8 & 56.2 & 61.2 & 52.8 & 43.7 & 57.0 \\
    \textbf{CRNN + ML} & \textbf{72.3} & \textbf{90.5} & \textbf{83.4} & \textbf{81.3} & \textbf{70.1} & \textbf{80.6} & \textbf{73.6} & \textbf{66.5} & \textbf{77.3} \\
    \bottomrule
    \end{tabular}
    
\end{table}

\begin{table}[h]
\centering
\caption{ADT Performance on DTM - $D_{D3}$ (\% f1-score)}
  \label{table_DD3_2}
\setlength{\tabcolsep}{3pt}  
 \begin{tabular}{@{}lccccccccc@{}}
    \toprule
    \textbf{Approach} & \textbf{KD} & \textbf{SD} & \textbf{HH} & \textbf{HH-O} & \textbf{TT} & \textbf{CY} & \textbf{RD} & \textbf{OT} & \textbf{Avg.} \\ 
    \midrule
    \textbf{SOTA-CRNN \cite{D7}} & 40.4 & 37.0 & 28.9 & 29.3 & 23.7 & 35.1 & 28.5 & 18.0 & 30.1 \\
    \textbf{PF-NMF \cite{R2}} & 41.3 & 39.7 & 35.6 & 33.8 & 30.4 & 36.9 & 32.6 & 24.9 & 34.4 \\
    \textbf{CRNN + TL} & 51.3 & 64.3 & 60.0 & 56.2 & 51.5 & 58.6 & 48.7 & 39.8 & 53.8 \\
    \textbf{CRNN + ML} & \textbf{70.0} & \textbf{86.1} & \textbf{78.5} & \textbf{78.2} & \textbf{67.3} & \textbf{76.9} & \textbf{63.6} & \textbf{60.1} & \textbf{72.6} \\
    \textbf{SS + SOTA-CRNN} & 44.8 & 40.1 & 33.0 & 31.5 & 26.8 & 36.3 & 29.4 & 21.1 & 32.9 \\
    \textbf{SS + PF-NMF} & 54.0 & 68.5 & 64.5 & 61.0 & 56.2 & 64.8 & 60.1 & 52.4 & 60.2 \\
    \textbf{SS + CRNN + TL} & 53.1 & 67.7 & 62.3 & 57.5 & 53.4 & 60.2 & 51.6 & 42.9 & 56.1 \\
    \textbf{SS + CRNN + ML} & \textbf{72.8} & \textbf{88.5} & \textbf{80.4} & \textbf{79.2} & \textbf{70.1} & \textbf{78.6} & \textbf{66.3} & \textbf{61.7} & \textbf{74.7} \\
    \bottomrule
    \end{tabular}

\begin{tablenotes}
    \item {\textbf{\textit{Note:}} Results use a 30ms collar}; SS: Source Separation using Demucs \cite{R1}
    \end{tablenotes}
\end{table}

\begin{table}[!h]
\centering
 \captionsetup{justification=centering}
\caption{Comparison of ADT Performance on DTM - $D_{D3}$ with Existing Methods}
\label{table_mdb_all}
\setlength{\tabcolsep}{6pt}  
\begin{tabular}{@{}lc@{}}
\toprule
\textbf{Approach} & \textbf{\% f1-score} \\ 
\midrule
\textbf{Proto-BL8-OFF \cite{F11}} & 67.0 \\
\textbf{Dense-BL8-OFF \cite{F11}} & 67.0 \\
\textbf{Proto-BL21-OFF \cite{F11}} & 69.0 \\
\textbf{Dense-BL21-OFF \cite{F11}} & 70.0 \\
\textbf{SOTA-CRNN \cite{D7}} & 30.1 \\
\textbf{SS + SOTA-CRNN} & 32.9 \\
\textbf{PF-NMF \cite{R2}} & 34.4 \\
\textbf{SS + PF-NMF} & 60.2 \\
\textbf{CRNN + TL} & 53.8 \\
\textbf{SS + CRNN + TL} & 56.1 \\ 
\textbf{CRNN + ML} & \textbf{72.6} \\
\textbf{SS + CRNN + ML} & \textbf{74.7} \\ 
\bottomrule
\end{tabular}
\begin{tablenotes}
    \item {\textbf{\textit{Note:}} Results use a 30ms collar and the same classes for fair comparison \\ \hspace{2cm} with \cite{F11}}; SS: Source Separation using Demucs \cite{R1}
    \end{tablenotes}
\end{table}

\section{Conclusion} \label{conclusion}

This paper proposes a novel Model-Agnostic Meta-Learning (MAML) based approach for Tabla Stroke Transcription (TST) and Automatic Drum Transcription (ADT), addressing challenges of limited annotated data and label heterogeneity. The method demonstrates effective transcription across diverse scenarios, including solo percussion, percussion with melodic accompaniments, and vocals, ensuring comprehensive validation and highlighting its robustness and adaptability.

Following TST, two novel $t\bar{a}la$ identification methods are introduced, advancing rhythmic analysis in Hindustani music. Experimental results show that the MAML-based approach consistently outperforms existing methods, even with limited labelled data.

Future work will focus on enhancing transcription accuracy for Indian music concert datasets with limited labelled data by leveraging advanced source separation techniques tailored to Indian music. We also aim to curate a larger and more diverse collection of polyphonic recordings to support training and evaluation in realistic conditions.

\bibliographystyle{IEEEtran}
\bibliography{main}

\end{document}